# Horizontal SCA Attacks against *kP* Algorithm Using K-Means and PCA

Marcin Aftowicz[1], Ievgen Kabin[1], Dan Klann[1], Yauhen Varabei[1], Zoya Dyka[1] and Peter Langendoerfer[1,2]

[1]*IHP – Leibniz-Institut für innovative Mikroelektronik,* Frankfurt (Oder), Germany
[2]*BTU Cottbus-Senftenberg,* Cottbus, Germany

*Abstract*— Side Channel Analysis attacks take advantage of the information leaked from the implementations of cryptographic algorithms. In this paper we describe two key revealing methods which are based on machine learning algorithms: K-means and PCA. We performed the attacks against ECDSA implementations without any prior knowledge about the key and achieved 100% accuracy for an implementation without any countermeasures against horizontal attacks and 88.7% accuracy for an implementation with bus address sequencing. In the scenario where the *kP* operation inputs are controlled by the attacker (as during signature verification), we achieved 98.3% accuracy for the implementation with countermeasures.

## I. INTRODUCTION

Elliptic Curve Cryptography (ECC) is widely used to guarantee the privacy and confidentiality for end users. Wireless sensor networks (WSN) and growing number of IoT devices relay on limited resources to provide their functionality. In such applications digital signature algorithms, providing data integrity and authenticity, need to guarantee low power consumption and low latency, yet with high level of security.

Elliptic Curve Digital Signature Algorithm (ECDSA) [1] is one of the algorithms recommended in Federal Information Processing Standard (FIPS) [2]. As long as the length of the key and type of the curve comply with the standard, the secret key cannot be revealed by analysing the inputs and outputs of the algorithm. However, every algorithm is implemented on a physical device, therefore it needs to be protected against Side Channel Analysis (SCA) attacks.

SCA attacks focus on measuring the physical effects like time, power consumption or electromagnetic radiation during execution of cryptographic operations. Since IoT devices and WSN can operate in large open areas, they are not protected against physical access. The attacker can measure and analyse a power trace (PT) or an electromagnetic trace (EMT) in order to reveal the secret i.e. the private key in case of ECC. If an attacker uses a single trace of a cryptographic operation to reveal the key, such attack is called a horizontal SCA attack. The most important operation to be analysed is the multiplication of an Elliptic Curve (EC) point *P* with a secret scalar *k*, called *kP* operation. In ECDSA an attacker can easy calculate the private key if the scalar *k* used for the signature generation is known to the attacker [3]. Thus, the goal of an attacker is to reveal the scalar *k*. In our previous paper [4] we have shown that a machine learning algorithm K-means is a successful method for revealing the secret scalar *k* attacking a single PT. We have noticed that another unsupervised method for dimensionality reduction and data visualization, called Principal Component Analysis (PCA), can also be used to extract the secret scalar *k* analysing a single measured *kP* trace. In this paper we evaluate the success of horizontal SCA attacks using PCA against three designs, described in [5]. We compare this new method with K-means method. We further use the knowledge resulting from K-means and PCA attacks and rerun the analysis only using the best leakage sources.

## II. IMPLEMENTATION DETAILS

Our design is a hardware accelerator for the EC point multiplication for the NIST EC *B-233* [2], i.e. it performs only a *kP* operation. The scalar *k* is an up to 233 bit long binary number and *P=(x, y)* is a point on the *B-233*. The EC point multiplication *kP* is the main operation in cryptographic protocols based on ECs. In ECDSA signature generation the scalar *k* is a random number and the EC point *P* is the base point *G* of the underlying elliptic curve. The coordinates of *G* are public and given in [2]. The theoretical basics of ECC are not the topic of this paper and are described in many books, for example in [6]. The Montgomery *kP* algorithm using Lopez-Dahab projective coordinates [7] is well-known and the most often implemented algorithm for the *kP* operation for ECs over *GF($2^n$)*. This algorithm is a bitwise processing of the scalar $k = k_{n-1} k_{n-2} \ldots k_1 k_0$, from its most significant bit $k_{n-1}$ down to its least significant bit $k_0$. The algorithm is fast and, as reported in literature [8, 9], resistant against simple SCA attacks, due to the fact that the sequence of operations – additions, squarings and multiplications of the field elements as well as write to register operations – doesn't depend on the processed bit value of the scalar *k*. The algorithm is often implemented in hardware as authentication chips or with the goal to accelerate the *kP* calculation to make the ECC approaches applicable for constrained devices.

In [5] we described the implementation details of our 3 different *kP* designs and evaluated their resistance against horizontal differential SCA attacks using statistical analysis. In [4] the K-means was evaluated as a mean for the extraction of the secret scalar *k* (further also denoted as key*)* using the same designs. In this paper we evaluate the applicability and effectiveness of Principal Component Analysis (PCA) as a mean for the secret revealing i.e. as an attacking method alternative to

K-means. We applied exactly the same steps for the preparation of traces as in [4] and [3].

### III. PREPERATION OF TRACES FOR SCA ATTACKS

#### A. Uncompressed Power Trace

In [4] and here we performed the horizontal attacks against an FPGA implementation of our $kP$ designs using the measured power trace of a single $kP$ execution. We ported our 3 ECC designs to a Spartan 6 FPGA [10] running at 4 MHz. We run the $kP$ execution and measured the current through the FPGA using a Riscure current probe [11]. We denote such trace in the rest of this paper as Power Trace (PT). For each of 3 designs one PT was captured during the $kP$ execution with the same inputs[1]. All 3 traces were captured using a LeCroy Waverunner 610Zi oscilloscope with a 2.5 GS/s sampling rate, i.e. with 625 measurement points – samples – per clock cycle.

Our implementation is based on Algorithm 2 in [12] that is a modification of the Montgomery $kP$ algorithm. In this modification the most and the second most significant bit, $k_{n-1}$ and $k_{n-2}$, are not processed in the main loop of the algorithm. The goal of this change is to prevent the revealing of $k_{n-2}$ with simple analysis and to increase the resistance of the implementation against template attacks. Thus in each PT we omit the analysis of the two most significant bits $k_{231}$ and $k_{230}$. We concentrate only on the revealing of $l = 230$ key bits $k_{229} \ldots k_1 k_0$ of our 232-bit long scalar $k$. Furthermore we introduce the following decomposition into:

1. $l = 230$ time **slots**, where each slot corresponds to the processing of a key bit $k_j$, where $j$ denotes the slot number $(l-1) \geq j \geq 0$ processed in the main loop of the algorithm.

2. Each slot consists of an equal number of $D = 54$ clock **cycles** and $i$ will denote the running number of the cycle within the slot, where $1 \leq i \leq D$.

3. Each clock cycle consists of $S = 625$ measured **samples**, with $s$ being the running number of the sample in the cycle and $1 \leq s \leq S$.

Each measured value within the main loop iteration, can be represented as the value of the $s^{th}$ sample in $i^{th}$ clock cycle in $j^{th}$ time slot as follows:

$$v_{j,i}^s \text{ where } \begin{aligned} & 1 \leq s \leq S; S = 625 \\ & 1 \leq i \leq D; D = 54 \\ & (l-1) \geq j \geq 0; l = 230 \end{aligned} \quad (1)$$

#### B. Compression of Traces – C3PT

Different compression methods for SCA attacks were proposed in the past [13, p. 86]. Depending on the selected compression method and signal-noise ratio in the measured trace compression can improve the success of attacks. In our experiments here and in [4] we applied the sum of squared values for representation of each clock cycle:

$$z_{j,i} = \sum_{s=1}^{S} (v_{j,i}^s)^2 \quad (2)$$

Here $z_{j,i}$ is a single value that represents a sum of squares of all samples in $i^{th}$ clock cycle in $j^{th}$ time slot in PT. We represent each clock cycle using only one value and obtain a Clock-Cycle-Compressed Power Trace (C3PT).

### IV. ANALYSIS METHODS

#### A. Data representation

Let us assume the following experiment: there are $l$ observations of some physical phenomenon. During each observation the phenomenon gets characterized by $d$ distinctive attributes $x$ (measured values). Such experiment could be summarized in form of an observation matrix:

$$X = \begin{bmatrix} x_{0,1} & \cdots & x_{0,d} \\ \vdots & \ddots & \vdots \\ x_{l-1,1} & \cdots & x_{l-1,d} \end{bmatrix} \quad (3)$$

In machine learning terminology, each attribute is called a feature and a single $j^{th}$ observation can be represented as a vector of its features:

$$x_j = (x_{j,1}, x_{j,2}, \ldots, x_{j,i}, \ldots, x_{j,d}) \quad (4)$$

In our case PTs and C3PTs can be summarized in form of observation matrices in different ways:

##### 1) Approach 1

We interpreted a PT as a set of $l = 230$ observations corresponding to slots. Every sample is treated as a single distinctive feature describing the slot. Each slot is represented as $d = S \cdot D = 33750$ samples. Using (1) this approach results in the following observation matrix:

$$X_1 = \begin{bmatrix} v_{1,1}^1 & \ldots & v_{1,1}^S & v_{1,2}^1 & \ldots & v_{1,2}^S & \ldots & v_{1,D}^S \\ & & & & & & \vdots & \vdots \\ v_{l-1,1}^1 & \ldots & v_{l-1,1}^S & v_{l-1,2}^1 & \ldots & v_{l-1,2}^S & \ldots & v_{l-1,D}^S \end{bmatrix} \quad (5)$$

We used the same approach to interpret a C3PT with $l = 230$ slots. Each slot consists of $D = 54$ clock cycles, hence there are $d = D = 54$ distinct features characterising each observation. Using (2) this approach can be represented as following observation matrix:

$$X_2 = \begin{bmatrix} z_{1,1} & \cdots & z_{1,D} \\ \vdots & & \vdots \\ z_{l-1,1} & \cdots & z_{l-1,D} \end{bmatrix} \quad (6)$$

---

[1] The hexadecimal representation of processed inputs, where $x$ and $y$ are affine coordinates of point $P$ on EC $B$ - $233$

$x$ = 181856adc1e7df1378491fa736f2d02e8acf1b9425eb2b061ff0e9e8246
$y$ = 9fed47b796480499cbaa86d8eb39457c49d5bf345a0757e46e2582de6
$k$ = 93919255fd4359f4c2b67dea456ef70a545a9c44d46f7f409f96cb52cc

*2) Approach 2*

In this approach a PT is represented in more than one matrix. In each of those matrices we used $l = 230$ slots as observations and a different clock cycle. In the $i^{th}$ experiment only $i^{th}$ clock cycle of every slot is used to characterise the entire slot. Each observation is described by $d = S = 625$ features reflecting the number of samples in the $i^{th}$ clock cycle. Using (1) this approach can be represented as $D = 54$ observation matrices, where the $i^{th}$ matrix takes the following form:

$$X_{3,i} = \begin{bmatrix} v_{1,i}^1 & \cdots & v_{1,i}^S \\ \vdots & & \vdots \\ v_{l-1,i}^1 & \cdots & v_{l-1,i}^S \end{bmatrix} \quad (7)$$

The same approach was used to interpret a C3PT. There are $D = 54$ matrices, each consist of $l = 230$ observations reflecting the number of slots in C3PT. Every slot is represented only by one clock cycle. Hence each observation is characterized by $d = 1$ feature. Using (2) this approach can be also represented as $D = 54$ observation matrices, where the $i^{th}$ matrix takes the following form:

$$X_{4,i} = \begin{bmatrix} z_{1,i} \\ \vdots \\ z_{l-1,i} \end{bmatrix} \quad (8)$$

*B. K-means clustering*

Due to the fact that the Montgomery $kP$ algorithm is a bitwise processing of the scalar $k$, the main idea of all statistical analysis methods is to distinguish the processing of a key bit '0' from a key bit '1'. The unsupervised iterative learning method *K-means* [14] uses adaptive grouping to divide a set of observations into $K$ independent classes (clusters, groups). It does so, whether any meaningful clustering is possible or not. The number of clusters $K$ and maximal number of iterations $M$ has to be determined manually prior to the analysis. *K-means* consists of the following steps:

1. Choose $K$ random observations as centroids (centres) of clusters, where $c_r$ is the $r^{th}$ cluster centroid:

$$c_r = (c_{r,1}, c_{r,2}, \ldots, c_{r,i}, \ldots, c_{r,d}) = random(x_j) \quad (9)$$

2. For all observations calculate the Euclidian distance $\rho_{j,r}$ between $j^{th}$ observation (4) and $r^{th}$ centroid.

$$\begin{aligned}
&\text{for each observation } x_j, \text{ where } (0 \leq j \leq l-1) \\
&\quad \text{for each centroid } c_r, \text{ where } (1 \leq r \leq K) \\
&\quad\quad \rho_{j,r} \leftarrow \sqrt{\sum_{i=1}^{d}(x_{j,i} - c_{r,i})^2}
\end{aligned} \quad (10)$$

3. Assign each observation to the cluster with the centroid laying closest to the observation. The distance vector $\boldsymbol{\rho}_j$ contains distances between $j^{th}$ observation and each of the K centroids.

$$\boldsymbol{\rho}_j = (\rho_{j,1}, \rho_{j,2}, \ldots, \rho_{j,r}, \ldots, \rho_{j,K}) \quad (11)$$

The index $r$ of the smallest element in $\boldsymbol{\rho}_j$ is the number of cluster $C_r$, where the $j^{th}$ observation gets assigned to.

$$\begin{aligned}
&\text{for each observation } x_j, \text{ where } (0 \leq j \leq l-1) \\
&\quad r \leftarrow index(min(\boldsymbol{\rho}_j)) \\
&\quad C_r.add\_observation(x_j)
\end{aligned} \quad (12)$$

4. Calculate the mean of all points assigned to one cluster. The mean becomes the new centroid.

$$\begin{aligned}
&\text{for each cluster } C_r, \text{ where } (1 \leq r \leq K) \\
&\quad \text{for each } j, \text{ where } x_j \in C_r \\
&\quad\quad c_r \leftarrow \frac{\sum_j x_j}{length(x_j)}
\end{aligned} \quad (13)$$

5. Jump back to step 2, unless:

   a. the clustering has not changed between the last two iterations, meaning that the clustering converged to an optimal solution (global or local minimum) and the centroids won't move anymore, or

   b. the maximal number of iterations $M$ has been reached. In this case the clustering has not converged and centroids are not arithmetic means of all points in the cluster.

Because the centroids in step 1 are chosen at random, the algorithm may not converge in a given number of iterations, or may converge to a local minimum. It is recommended to run the algorithm more than once and choose the converged solution (where centroids are means of clustered points) and where the sum of distances between all clustered points and their cluster centroid is minimal.

K-Means is part of an open source scikit-learn library (v0.20.2). All traces were analysed using Python (v3.6.8) in the Anaconda (v1.9.6) environment. The results of the analysis are given in Section V.

*C. Pricipal Component Analysis*

High dimensionality of the data makes it impossible to plot the entire dataset in 'usual' coordinate system, i.e. 3D. In our experiments, the dimensions span between $d = 1$ (see formula (8)) and $d = 33750$ (see formula (5)). Moreover some features are redundant to the analysis, i.e. current flow during some clock cycles is independent of the processed key bit. Unsupervised learning contains a set of tools, called dimensionality reduction algorithms, which were designed to i.a. visualize such data. The Principal Component Analysis [15] helps to focus on dimensions, which are uncorrelated and have the biggest variance, assuming that features, which don't change a lot are not as important as those with significant variance. Please note that this assumption is not always true.

The first stage of the PCA is to linearly transform a *d*-dimensional dataset into a new *d*-dimensional coordinate system, such that all dimensions are uncorrelated and sorted according to their variance. This new set of variables is called the principal components. The principal components are calculated using either the correlation or the covariance matrix of the dataset. The reduction of dimensions happens in the later stage, by getting rid of the components with the smallest variance.

*1) Covariance and correlation of two random variables*

The covariance of two variables indicates if those variables are correlated. Positive covariance means that when one variable grows, so does the other. Negative – the opposite. When covariance is zero, two variables are not correlated. The correlation is a dimensionless representation of covariance (the same relationship, but scaled between -1 and 1). Maximal positive or negative correlation, denoted as $\pm 1$, means that one variable is a linear function of another. The correlation and covariance of two variables are bound together with the following formula:

$$corr(\alpha, \beta) = \frac{cov(\alpha, \beta)}{\sigma_\alpha \sigma_\beta} \qquad (14)$$

where $\sigma$ is the standard deviation of the random variable.

*2) Correlation and covariance matrix*

In PCA the random variable is a feature vector of the observation matrix. Let $X$ be the observation matrix with $l$ observations and $d$ features as shown in (3). Let $\pmb{x}_j$ be the $j^{th}$ observation vector with $d$ features as described in (4). Further let us define $\pmb{x}_i$ to be an $i^{th}$ feature vector with $l$ observations of the following form:

$$\pmb{x}_i = (x_{0,i}, x_{1,i}, \dots, x_{j,i}, \dots, x_{l-1,i})^T \qquad (15)$$

The correlation matrix looks as follows:

$$\sum = \begin{bmatrix} corr(\pmb{x}_1, \pmb{x}_1) & \cdots & corr(\pmb{x}_1, \pmb{x}_d) \\ \vdots & \ddots & \vdots \\ corr(\pmb{x}_d, \pmb{x}_1) & \cdots & corr(\pmb{x}_d, \pmb{x}_d) \end{bmatrix} \qquad (16)$$

Correlation matrix $\sum$ is a square symmetric matrix. All terms laying on the main diagonal of a $\sum$ represent the correlation of each variable with itself, i.e. they are all equal +1. All other terms represent the correlation between two different pairs of feature vectors. Any two highly correlated feature vectors will be visible in $\sum$ (values close to $\pm 1$).

The covariance matrix is also a square symmetric matrix, where terms on the main diagonal simply express the variance (squared standard deviation) of each feature vector and all other terms – the covariance between pairs of feature vectors.

*3) When to use the covariance, when the correlation matrix*

According to (14) the correlation matrix is a scaled, dimensionless version of covariance, where the scaling factor is the standard deviation of both feature vectors. Therefore the choice between using the correlation or the covariance matrix, as a base for PCA, depends whether all variables lay on the same scale or not. If all measured features represent the same physical quantity measured in the same units, the covariance matrix is the right choice. If features lay on different scales or represent different quantities, or some feature vectors have much higher variance than others – the correlation matrix should be used [15].

It was shown in [16] and [17] that after Side Channel Analysis, the clock cycles which showed the biggest amplitude changes between slots (biggest variance in feature vector) could, but not necessarily did, serve as best leakage sources. Therefore, we have standardized the dataset prior to the analysis (17). The standardized feature vector $\pmb{x}_i^*$ can be computed as follows:

$$\pmb{x}_i^* = \pmb{x}_i - \frac{\mu_i}{\sigma_i}\pmb{1} \qquad (17)$$

where $\mu_i$ is the arithmetic mean of $\pmb{x}_i$ and $\sigma_i$ is the standard deviation of $\pmb{x}_i$ and $\pmb{1}$ is a column vector of ones of length $d$. The standardization removes the mean (centred dataset) and scales the vector by standard deviation. Let $X^*$ be called the standardized observation matrix, such that all feature vectors in $X^*$ have mean equal to zero and standard deviation equal to one.

According to (14), the standardisation of feature vectors before calculating the covariance matrix is equivalent to calculating the correlation matrix of original feature vectors (centring the dataset doesn't influence the covariance calculation [15]) . The matrix $\sum$ from (16) is equivalent to:

$$\sum = \begin{bmatrix} cov(\pmb{x}_1^*, \pmb{x}_1^*) & \cdots & cov(\pmb{x}_1^*, \pmb{x}_d^*) \\ \vdots & \ddots & \vdots \\ cov(\pmb{x}_d^*, \pmb{x}_1^*) & \cdots & cov(\pmb{x}_d^*, \pmb{x}_d^*) \end{bmatrix} \qquad (18)$$

Despite the fact that all features in our experiments are on the same scale (compressed values in the C3PT and measured current flow in the PT), we used the covariance matrix of standardized dataset, which is equivalent to using the correlation matrix. In this way, the features with high variance won't dominate the outcome of the PCA.

*4) Eigendecomposition of the covariance matrix*

In order to linearly transform the observation matrix into the principal component space some transformation matrix needs to be applied. This transformation matrix is found during the process of eigendecomposition of the covariance matrix $\sum$. A square, non-defective covariance matrix, such as $\sum$, can be represented as eigenvectors and eigenvalues. The search for eigenvectors and eigenvalues is not a subject of this paper, but there are some important properties worth mentioning, to give intuition towards how PCA works.

Eigenvectors of the covariance matrix $\sum$ are orthonormal (perpendicular to each other) unit vectors. They express the linear combination of original coordinates of the observation matrix $X$. There is an eigenvalue associated with each eigenvector. The eigenvalue represents the variance of this linear combination, i.e. the eigenvector with the biggest eigenvalue, shows the direction, in the original observation space $X$, with the biggest variance. This linear combination shows the direction of the first principal component in the original observation.

For a $d$-dimensional squared covariance matrix $\sum$ there exist $d$ eigenvectors $v_i$, which are orthonormal unit vectors and therefore can form a new coordinate system. The vectors are sorted in descending order based on their eigenvalues $\lambda_i$. The observation matrix $X$ gets transformed into $\hat{X}$ by:

$$X \begin{bmatrix} v_1 \\ \vdots \\ v_d \end{bmatrix} = \hat{X} \qquad (19)$$

It means that first component represents a new dimension, along which the transformed dataset $\hat{X}$ has the biggest variance (widest span). Every next component will represent (explain) smaller and smaller part of the variance. The dimensionality reduction consists in choosing only the first components of $\hat{X}$, preserving just a part of original variance.

## 5) PCA graphical representation

Each experiment (summarised as an observation matrix) could also be represented as a group of points in a coordinate system. If an observation can be treated as a vector of its $d$ attributes, then each of the $l$ observations can be represented as a single point in a $d$-dimensional space (coordinate system with $d$ coordinates). In terms of graphical representation, the standardization step (17) centres the data on the mean value of each dimension, making values bigger from the mean to be positive, and smaller from the mean – negative. It moves the origin of the coordinate system into the mean point of each features. The division by standard deviation scales every dimension. The transformation with eigenvectors (19) rotates the coordinate system so that the first axis shows the direction of the biggest variance, and the second coordinate – the second biggest variance, and so on.

The visualization in Figure 1 (left) shows how PCA finds the principal components in a $d = 3$ dimensional space. Each out of $l = 230$ points represents a single slot. The $d = 3$ features are the chosen compressed clock cycle values with highest leakage. The first component takes the direction, along which the data set has the biggest variance (black vector – PC1). Two other components are orthogonal to the first one and all together form the new coordinate system. The dimensionality reduction, from 3D to 2D, consist in eliminating the third principal component (green vector – PC3). Figure 1 (right) represents the three dimensional observation matrix in the space of its first two principal components

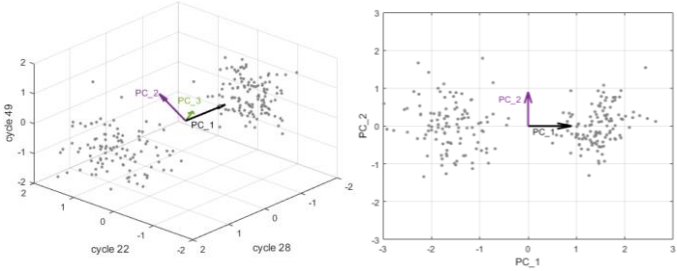

Figure 1. Visualization of an experiment with 230 observations, in our case slots, plotted in the space of their 3 features - compressed values of clock cycles (left) and in the space of their two first principal components (right)

## 6) Classification method based on PCA

In [4] we used PCA to visualize the attack results. We noticed that the points formed two dense patterns, as visible in Figure 1 (right). The classification can be performed based on the first principal component:

- points with positive value of PC_1 belong to class1
- points with negative value of PC_1 belong to class2

The assumption is that points belonging to class1 represent processing of different scalar bits than point in class2. This assumption was used to extract the key. The results are summarized in Section V.

The implementation of PCA used in this paper is a part of an open source scikit-learn (v0.20.2) library in Anaconda (v1.9.6) Python (v3.6.8) environment.

## V. ATTACK RESULTS

We conducted SCA attacks on all 3 Designs using different approaches in interpreting their PTs and C3PTs. We used K-means clustering method (Section IV.B) with $K = 2$ centroids ('0' or '1' as key bit value), $M = 300$ maximal number of iterations and 10 repetitions with always newly randomized centroids. We also used PCA (Section IV.C), classifying points based on the first principal component.

### 1) Attack1

We used *Approach 1* (see section IV.A.1) to interpret the compressed and uncompressed PTs of all 3 Designs. Each trace represents a single experiment summarised as observation matrices (5) and (6). All feature vectors in observation matrices (5) and (6) were standardized according to (17) prior to applying any key revealing means. We used K-means clustering and PCA classification on the standardized observation matrices and obtained one key candidate per attacked trace.

In order to evaluate our analysis we compared the key candidates with the real key that we know as a designer. Out of $l = 230$ key bits that were analysed some were guessed correctly and some not. Both, K-means and PCA, can extract two classes, but they cannot tell which class represents the ones and which – the zeroes. We can always switch the classes and invert the result. Therefore the success rate of the attack is expressed as relative correctness. The relative correctness falls between 50% (the worst outcome, where only half of the bits were guessed correctly) and 100% (where all bits were guessed correctly, or all bits were guessed wrongly and the opposite assumption is correct). The results are summarized in Table 1.

TABLE 1. RESULTS OF ATTACK1 IN FORM OF RELATIVE CORRECTNESS

| analysis method | Design 1 | | Design 2 | | Design 3 | |
| --- | --- | --- | --- | --- | --- | --- |
| | PT | C3PT | PT | C3PT | PT | C3PT |
| K-means | 100% | 100% | 53.9% | 100% | 51.3% | 57.4% |
| PCA | 54.8% | 100% | 53.5% | 100% | 51.7% | 56.5% |

#### a) Graphical representation of chosen results

Figure 2 visualizes the results of *Attack1* on compressed PT of Design1. There are 3 plots. Each plot contains $l = 230$ points representing observations, plotted in the space of their first two principal components. Each point represents a single slot and is associated with processing of a single bit of secret scalar $k$. In the left plot, points are coloured based on the designers' knowledge about the processed key bit values. Points marked light blue, represent processing of zeros and points marked dark blue – ones. The middle plot shows the result of K-means clustering. Blue and red colours represent two classes. All slots were clustered correctly. The right plot shows the result of PCA classification, where all slots were classified correctly as well.

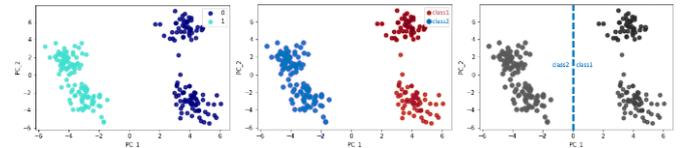

Figure 2. Result of *Attack1* using compressed PT of Design1. Designers' knowledge (left). K-means clustering – 100% (middle). PCA classification – 100% (right)

The biggest variance of the dataset shown in Figure 2, results from the big Euclidian distance between clusters. There are even 4 clusters to be seen in Figure 2. They correspond to differences in processing consecutive key bits, i.e. the processing of bit '0' differs if the previously processed bit was '1' or '0'. The same applies to processing of bit '1'.

Figure 3 represents the results of *Attack1* on uncompressed PT of Design1. The middle plot shows that the K-means correctly clustered the data. Classification with PCA, visible in the right plot, shows the correct classification is given not by the first, but by the second principal component. It turns out that the cluster inner variance was bigger than the variance caused by processing of different key bits. The variance resulting from processing different key bits, was the second biggest contribution to the entire variance of the dataset, hence it is visible in the second principal component.

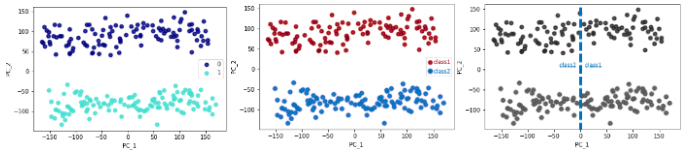

Figure 3. Result of *Attack1* using PT of Design1.
Designers' knowledge (left). K-means clustering – 100% (middle).
PCA classification – 54.8% (right).

The biggest variance in C3PT of Design3 (see Figure 4), explained by the first principal component has nothing to do with the processing of different key bits. There is another factor influencing the results of PCA analysis. The artificial noise created by sequential change of addresses of the design blocks and the data on the bus causes higher variance than the differences between processing of different key bits. Both, the K-means and the PCA got the same – wrong result (they differ in classification of only one bit).

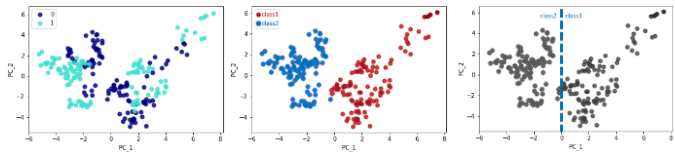

Figure 4. Result of *Attack1* using compressed PT of Design3.
Designers' knowledge (left). K-means clustering – 51.3% (middle).
PCA classification – 51.7% (right)

*2) Attack2*

We used *Approach 2* (see section IV-A) to interpret the compressed and uncompressed traces of all 3 Designs. Each trace represents $D = 54$ experiments, each consisting of $l = 230$ observations with $d = S = 625$ attributes for uncompressed PT and $d = 1$ attribute for the compressed PT. Each experiment uses single clock cycle to reveal the key. All feature vectors for all experiments, represented as observation matrices (7) and (8) were standardized, according to (17), prior to applying any key revealing methods. We used K-means clustering and PCA classification on the standardized observation matrices and obtained one key candidate per experiment, resulting in $D = 54$ key candidates per attacked trace.

We compared each of the extracted key candidates with the real key to evaluate the success rate of the attack. Not only the high correctness of the best key candidates, but also the number of the key candidates with a high correctness, can characterize the success of the attacks. Table 2 shows the results of *Attack2*, i.e. the highest relative correctness δ achieved during all 54 experiments in range 50% – 100% and the number of key candidates #$k_{cand}$ with δ>95%.

TABLE 2. RESULTS OF ATTACK2

| analysis method and evaluation criteria | | Design 1 | | Design 2 | | Design 3 | |
|---|---|---|---|---|---|---|---|
| | | PT | C3PT | PT | C3PT | PT | C3PT |
| K-means | δ | 97.4% | 99.1% | 100% | 99.6% | 68.7% | 88.3% |
| | #$k_{cand}$ (δ>95%) | 2 | 8 | 3 | 5 | 0 | 0 |
| PCA | δ | 97.4% | 99.6% | 100% | 100% | 69.1% | 88.7% |
| | #$k_{cand}$ (δ>95%) | 2 | 8 | 2 | 5 | 0 | 0 |

Please note that we conducted *Attack1* and *Attack2* without any prior knowledge about the processed scalar.

*3) Attack3 as a combination of Attack2 and Attack1*

We have showed in *Attack1* and *Attack2* that Design1 and Design2 are not resistant against the K-means and PCA attacks. Attacks on Design3 on the other hand, resulted in only 88.7% success rate in the best case (204 out of 230 bits revealed). We combined *Attack1 and Attack2* to aim for a higher success rate.

$kP$ operation is the main operation in signature generation and verification algorithms. In signature generation, the scalar is hidden from the attacker, but in signature verification, the attacker can choose its value freely. If the same design is used to accelerate both signature generation and verification, the attacker can measure the power trace of the $kP$ operation during signature verification algorithm. Later they can apply *Attack2* and evaluate its success rate, since they know the scalar. In this way they reveal the clock cycles, which are good leakage sources.

In *Attack3* we used *Approach 1* to interpret the compressed and uncompressed PTs of all 3 Designs (as during *Attack1*). Additionally we used the knowledge about good leakage sources from *Attack2*, i.e. we can sort all clock cycles according to the strength of their leakage, based on *Attack2*. We designed $D = 54$ experiments. The first experiment is equivalent to *Attack1*, i.e. we use all clock cycles of a trace to find one key candidate. The second experiment eliminates the clock cycle with the lowest leakage, i.e. $S = 625$ features in PT and one feature in compressed PT are removed from the observation matrix. Every next experiment eliminates another clock cycle with the lowest leakage, until the last experiment, when only one clock cycle is left. This last experiment is equivalent to the experiment in *Attack2* which revealed the strongest leakage source. Each experiment uses decreasing number of features and results in one key candidate.

We have compared all 54 key candidates with the real key. We calculated the relative correctness of each candidate. In Table 3 we presented the relative correctness δ of the best key candidate and the minimal number η of clock cycles (features) used in the experiment.

TABLE 3. RESULTS OF ATTACK3

| analysis method and evaluation criteria | | Design 1 | | Design 2 | | Design 3 | |
|---|---|---|---|---|---|---|---|
| | | PT | C3PT | PT | C3PT | PT | C3PT |
| K-means | δ | 100% | 100% | 100% | 100% | 70.4% | 98.3% |
| | η | 2 | 2 | 1 | 2 | 2 | 6 |
| PCA | δ | 100% | 100% | 100% | 100% | 71.7% | 96.1% |
| | η | 2 | 2 | 1 | 1 | 5 | 4 |

*a) graphical representation of chosen results*

Figure 5, similarly to Figure 2 – Figure 3, represents the results of *Attack3* using compressed PT of Design3.

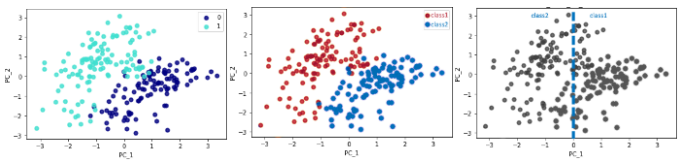

Figure 5. Result of *Attack3* using C3PT of Design3 for 6 clock cycles with strongest leakage. Designers' knowledge (left).
K-means clustering – 98.3% (middle). PCA classification – 96.1% (right).

The combination of clock cycles with strongest leakage resulted in best clustering and classification. K-means outperformed PCA in guessing several bits, but both methods showed strong improvement in *Attack3*. Feature selection combined with clock cycle compression allowed to emphasize the differences in processing of different scalar bits and reduce the influence of artificial noise, injected via countermeasures proposed in [3].

## VI. CONCLUSIONS

Both applied here analysis methods – K-means and PCA – performed better when analysing the compressed PTs, because the useful information resides only in part of the clock cycle (activity on rising edge of the clock) and compression reduces the level of noise. K-means clustering turned out to have comparable results with the PCA classification; with only one exception – the result of Attack1 against Design1 using uncompressed power trace. None of the presented methods was efficient in extracting the key information from Design3 and only the knowledge about clock cycles with highest leakage, allowed to achieve the correctness of the revealed key above 95%. The unsupervised machine learning methods are valuable tools in analysing multidimensional datasets, but with such little number of observations ($l = 230$) they were not able to distinguish between the key dependant activity and artificially injected changes in the bus addressing scheme. *Attack3* requires an additional trace of the *kP* execution with a known scalar *k*. Such trace can be captured during signature verification. Opposite to supervised learning methods we don't perform the typical training. But we used this additional measured *kP* trace to reduce the number of features. It allows to make better decisions using unsupervised machine learning algorithms.